\documentclass[apj,]{emulateapj}
\usepackage{amsmath}
\usepackage{amssymb}
\usepackage{euscript}
\usepackage{algorithm2e,algorithmic}
\usepackage{bbold}
\usepackage{paralist}
\usepackage{xcolor}

\def\gtorder{\mathrel{\raise.3ex\hbox{$>$}\mkern-14mu
             \lower0.6ex\hbox{$\sim$}}}
\def\ltorder{\mathrel{\raise.3ex\hbox{$<$}\mkern-14mu
             \lower0.6ex\hbox{$\sim$}}}

\slugcomment{Draft of \today}

\shortauthors{Soumagnac et al.}

\begin{document}


\title{Supernova PTF\,12glz: a possible shock breakout driven through an aspherical wind}



\author{Maayane T. Soumagnac\altaffilmark{1}, Eran O. Ofek\altaffilmark{1}, Avishay Gal-yam\altaffilmark{1}, Eli Waxman \altaffilmark{1}, Sivan Ginzburg \altaffilmark{2}, Nora Linn Strotjohann\altaffilmark{3}, Tom A. Barlow\altaffilmark{4}, Ehud Behar\altaffilmark{5}, Doron Chelouche\altaffilmark{6}, Christoffer Fremling \altaffilmark{7}, Noam Ganot \altaffilmark{1}, Suvi Gezari\altaffilmark{8}, Mansi M. Kasliwal \altaffilmark{9}, Shai Kaspi\altaffilmark{12}, Shrinivas R. Kulkarni \altaffilmark{9}, Russ R. Laher\altaffilmark{11}, Dan Maoz\altaffilmark{12}, Christopher D. Martin\altaffilmark{4}, Ehud Nakar\altaffilmark{12}, James D. Neill\altaffilmark{4}, Peter E. Nugent \altaffilmark{13,14}, Dovi Poznanski\altaffilmark{12}, Steve Schulze\altaffilmark{1}, Ofer Yaron\altaffilmark{1}}

\altaffiltext{1}{Department of Particle Physics and Astrophysics, Weizmann Institute of Science, Rehovot 76100, Israel.}
\altaffiltext{2}{Racah Institute of Physics, The Hebrew University, Jerusalem 91904, Israel.}
\altaffiltext{3}{Desy Zeuthen, 15738 Zeuthen, Germany.}
\altaffiltext{4}{California Institute of Technology, 1200 East California Boulevard, MC 278-17, Pasadena, CA 91125, USA.}
\altaffiltext{5}{Physics Department, Technion Israel Institute of Technology, 32000 Haifa, Israel.}
\altaffiltext{6}{Haifa Center for Theoretical Physics and Astrophysics, Faculty of Natural Sciences, University of Haifa, Haifa 3498838, Israel}
\altaffiltext{7}{Department of Astronomy, The Oskar Klein Center, Stockholm University, AlbaNova, 10691 Stockholm, Sweden.}
\altaffiltext{8}{Department of Astronomy, University of Maryland, College Park, MD 20742-2421, USA.}
\altaffiltext{9}{Cahill Center for Astrophysics, California Institute of Technology, Pasadena, CA, 91125, USA.}
\altaffiltext{10}{Spitzer Science Center, California Institute of Technology, Pasadena, CA 91125.}
\altaffiltext{11}{Infrared Processing and Analysis Center, California Institute of Technology, Pasadena, CA 91125, U.S.A.}
\altaffiltext{12}{School of Physics and Astronomy, Tel Aviv University, 69978 Tel Aviv, Israel.}
\altaffiltext{13}{Lawrence Berkeley National Laboratory, Berkeley, CA, 94720, USA.}
\altaffiltext{14}{Department of Astronomy, University of California, Berkeley, CA, 94720-3411, USA.}







\begin{abstract}
We present visible-light and ultraviolet ($UV$) observations of the supernova PTF\,12glz. The SN was discovered and monitored in near-$UV$ and R bands as part of a joint \textit{GALEX} and Palomar Transient Factory campaign. It is among the most energetic Type IIn supernovae observed to date ($\approx10^{51}$ erg). If the radiated energy mainly came from the thermalization of the shock kinetic energy, we show that PTF\,12glz was surrounded by $\sim1\,\rm{M}_{\odot}$ of circumstellar material (CSM) prior to its explosive death. PTF\,12glz shows a puzzling peculiarity: at early times, while the freely expanding ejecta are presumably masked by the optically thick CSM, the radius of the blackbody that best fits the observations grows at $\approx7000$\,km\,s$^{-1}$. Such a velocity is characteristic of fast moving ejecta rather than optically thick CSM. This phase of radial expansion takes place before any spectroscopic signature of expanding ejecta appears in the spectrum and while both the spectroscopic data and the bolometric luminosity seem to indicate that the CSM is optically thick. We propose a geometrical solution to this puzzle, involving an aspherical structure of the CSM around PTF\,12glz. By modelling radiative diffusion through a slab of CSM, we show that an aspherical geometry of the CSM can result in a growing effective radius. This simple model also allows us to recover the decreasing blackbody temperature of  PTF\,12glz. \texttt{SLAB-Diffusion}, the code we wrote to model the radiative diffusion of photons through a slab of CSM and evaluate the observed radius and temperature, is made available on-line.
\end{abstract}



\section{Introduction}

Type IIn supernovae (SNe) are characterized by prominent and narrow-to-intermediate width Balmer emission lines in their spectra \citep{Schlegel1990, Filippenko1997, Smith2014,Gal-Yam2016}. Rather than a signature of the explosion itself, this spectral specificity is presumably the result of the photoionization of a dense, Hydrogen-rich, circumstellar medium (CSM) which is ejected from the SN progenitor prior to the explosion. 

The Type IIn class is not a well-defined category of objects, as many SNe show the characteristic narrow Balmer lines in their spectra, sometime during their evolution. These lines are the signature of an external physical phenomenon highly dependent on the surrounding environment, rather than of any intrinsic property of the explosion. Depending on the spatial distribution and physical properties of the CSM, these lines may persist for days (``flash spectroscopy'', \citealt{Gal-Yam2014,Khazov2016,Yaron2017}), weeks (e.g., SN\,1998s, \citealt{Li1998, Fassia2000, Fassia2001}; SN\,2005gl, \citealt{Gal-Yam2007}; SN\,2010mc, \citealt{Ofek2013}), or years (e.g., SN\,1988Z, \citealt{Danziger1991,Stathakis1991,Turatto1993,VanDyk1993,Chugai1994,Fabian1996,Aretxaga1999,Williams2002,Schlegel2006,Smith2017}; 2010\,jl, \citealt{Patat2011, Stoll2011, Gall2014,Ofek2014}).




In the last decades, the physical picture governing SN IIn explosions and the wider family of ``interacting'' SNe - SNe whose radiation can be partially or completely accounted for by the ejecta crashing into a dense surrounding medium - has become clearer (see e.g., \citealt{Chevalier1982}, \citealt{Chugai1994}, \citealt{Chugai2004}, \citealt{Ofek2010}, \citealt{Chevalier2011}, \citealt{Ginzburg2014}, \citealt{Moriya2014}). In recent years, there is growing evidence that, in the majority of cases, the high-density CSM originates from explosive phenomena taking place in the months to years prior to the SN explosion. One piece of evidence supporting this conclusion is the direct detection of the so-called precursors (luminous outbursts) in the months to years prior to the SN explosion (e.g., \citealt{Foley2007, Pastorello2007, Fraser2013, Ofek2013, Ofek2014c, Ofek2016, Elias-Rosa2016, Thone2017}). Several theoretical mechanisms have been suggested to explain extreme mass-loss episodes in the final stages of stellar evolution (e.g., \citealt{Woosley2007,Quataert2012,Chevalier2012b,Soker2016}). 

While in normal core-collapse SNe, the radiation-mediated shock breaks out upon reaching the stellar surface, producing a strong blast in the $UV$ and X-rays \citep{Nakar2010,Rabinak2011}, in the case of SNe IIn the ejecta may crash into the optically thick CSM. The radiation-dominated and radiation-mediated shock runs into the CSM surrounding the star and goes on propagating into it as long as $\tau \gtrapprox c/v_{sh}$, where $\tau$ is the optical depth from the shock to the edge of the wind, $v_{sh}$ is the shock velocity, and $c$ is the speed of light (e.g., \citealt{Ofek2010}). When $\tau \sim c/v_{sh}$, (this condition is verified when the timescale for photons to diffuse from the shocked region to the photosphere becomes comparable to the dynamical timescale of the shock), the shock breaks out: photons diffuse ahead of the shock faster than the ejecta and radiation can escape ahead of the shock \citep{Weaver1976}. After the shock breakout, in the presence of massive CSM above the shock, the radiation-dominated shock transforms into a collisionless shock \citep{Katz2011,Murase2011,Murase2014}. The collisionless shock slows down the ejecta and converts its kinetic energy into hard X-ray photons \citep{Katz2011,Murase2011,Murase2014}. If the optical depth of the CSM above the shock is high enough, the X-rays generated in the collisionless shock are converted into $UV$ and visible radiation (e.g., \citealt{Chevalier2012, Svirski2012}). Without a sufficient optical depth though, the bulk of the X-ray photons will not convert into optical photons. 


As far as a spectral signature is concerned, the common picture explaining SNe IIn observations is as follows. As long as the CSM is optically thick, the photosphere which emits the continuum is located in the unshocked CSM, masking the observer's view of the shock. The radiation from the shock propagates upstream and photoionizes the slowly moving CSM, resulting in relatively narrow Balmer recombination emission lines in the SN spectrum. As the shock reaches the optically thin medium, broader components can appear in the spectrum - maybe arising from the shocked zone forming at the contact discontinuity between the decelerated ejecta and the shocked CSM \citep{Chugai2004}. Alternatively, if the CSM is optically thin, lines from the fast moving SN ejecta, generated in inner regions, may become visible (e.g., \citealt{Chevalier1994}).

Observing SNe IIn at wavelengths where the collisionless shock radiates most - namely  $UV$ and X-rays - has the potential to unveil precious information about the explosion mechanism and the CSM properties (e.g., \citealt{Ofek2013a}). In particular, it may provide a much better estimate of the bolometric luminosity of the event. 
In this paper, we present and analyse the $UV$ and visible-light observations of PTF\,12glz, a SN IIn observed in a joint campaign by \textit{GALEX} and the Palomar Transient Factory (PTF) and detected in the $UV$. PTF\,12glz is one of the six SNe discovered during this campaign \citep{Ganot2016}. The survey was carried out as a proof-of-concept for the $ULTRASAT$ mission \citep{Sagiv2014}.
 
Observations of SNe IIn are usually analyzed within the framework of spherically symmetric models of CSM. However, resolved images of stars undergoing considerable mass loss (e.g., $\eta$ Carinae; \citealt{Davidson1997, Davidson2012}), as well as polarimetry observations \citep{Leonard2000, Hoffman2008, Wang2008, Reilly2017} suggest that asphericity should be taken into account for more realistic modeling. Asphericity of the CSM has recently been invoked to interpret the spectrocopic and spectropolarimetric observations of the Type IIn SN SN\,2012ab \citep{Bilinski2017} and SN\,2009ip \citep{Mauerhan2014,Smith2014_2009ip,Levesque2014,Reilly2017}. In this paper, we show that the light curve of PTF\,12glz may be interpreted as evidence for aspherical CSM.
 
We present the aforementioned observations of PTF\,12glz in \S 2. In \S 3, we present the analysis of these observations and the puzzling inconsistency between the spectroscopic and photometric observations. In \S 4, we model the radiative diffusion of photons through an aspherical slab and propose a solution to this puzzle. We then summarize our main results in \S 5. In the Appendix, we make available \texttt{SLAB-Diffusion}\citep[][Codebase: \url{https://github.com/maayane/SLAB-Diffusion}]{SLAB-Diffusion}, a computer code for modeling radiation through a slab of CSM.

\section{Observations and data reduction}
In this section, we present the observations of PTF\,12glz by the \textit{GALEX}/PTF $UV$ wide-field transient survey. This campaign, conducted during a nine-week period from 2012 May 24 through 2012 July 28, used the \textit{GALEX} $NUV$ camera to cover a total area of about $600$ deg$^2$ over 20 times with a three-day cadence, while PTF observed the same region with a two-day cadence \citep{Ganot2016}. 

\subsection{Discovery}\label{sec:discovery}
\begin{figure}
\begin{center}
\includegraphics[scale=.35]{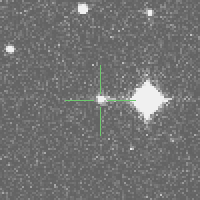}
\includegraphics[scale=.35]{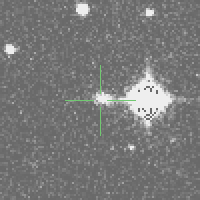}
\includegraphics[scale=.35]{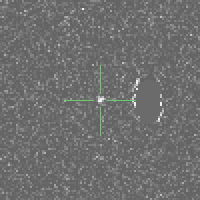}
\includegraphics[scale=.40]{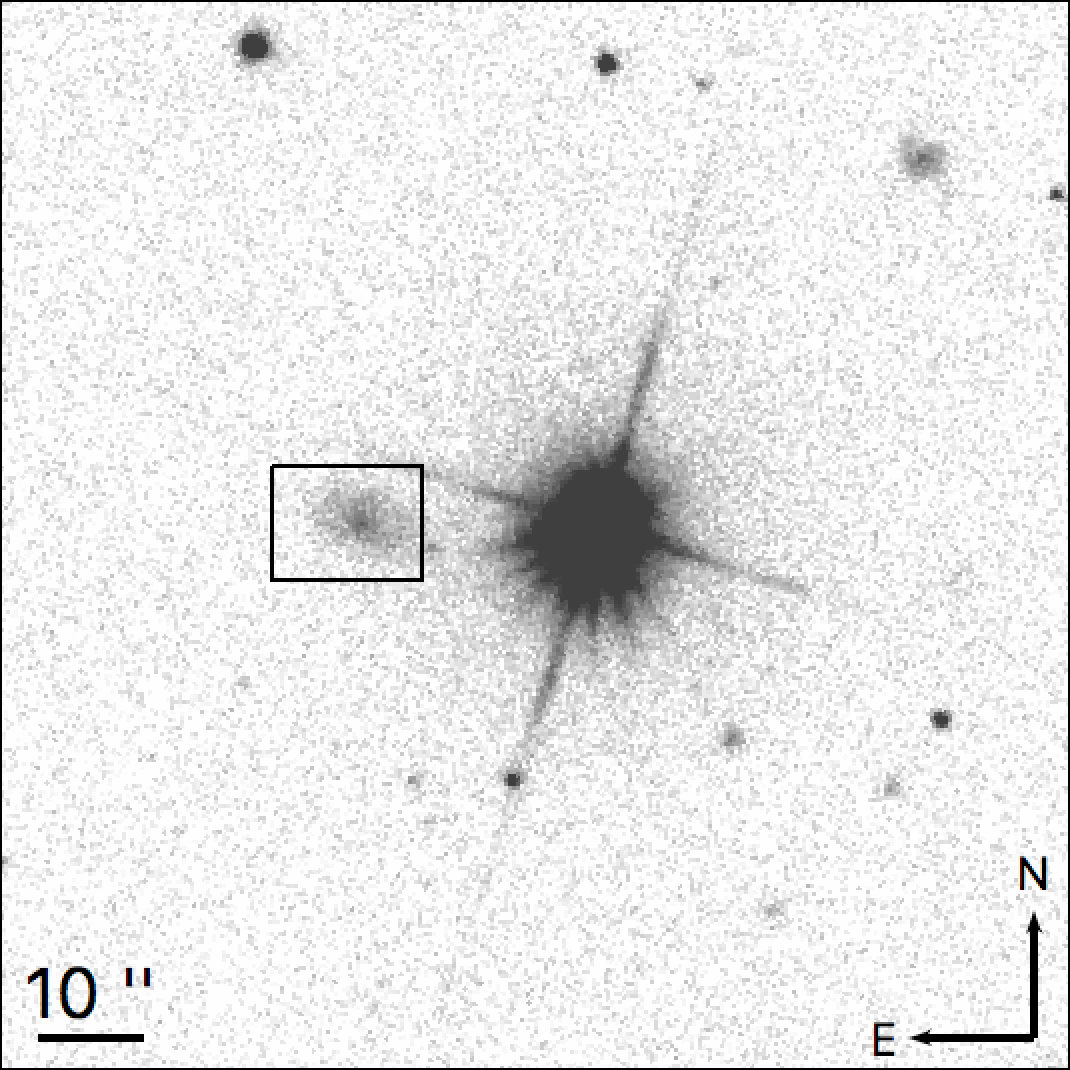}
\caption{Top panels: (left to right) the discovery image, reference image, and subtracted P48 image of PTF\,12glz. Lower panel: the SDSS image of J155452.95+033207.5, the host of the supernova PTF\,12glz. 
The box encircles the host: $\alpha=238.72100^\circ$ and $\delta=3.53542^\circ$. Credit: SDSS.}\label{fig:host}
\label{fig:lc}
\end{center}
\end{figure}

PTF\,12glz was discovered on 2012 July 7 by the PTF \citep{Law2009,Rau2009} automatic pipeline reviewing potential transients in the data from the PTF camera mounted on the $1.2$\,m Samuel Oschin telescope (P48, \citealt{Rahmer2008}). The image processing pipeline is discussed in \cite{Laher2014} and the photometric calibration is described in \cite{Ofek2012}. The SN is associated with an $r=18.51$ mag galaxy, SDSS\footnote{Sloan Digital Sky Survey; \cite{York2000}}J155452.95+033207.5, shown in Figure~\ref{fig:host} and modelled in section~\ref{sec:host}. The coordinates of the object, measured in the PTF images are $\alpha=15^h54^m53^s.04$, $\delta=+03^d32'07''.5$ (J2000.0). The redshift $z=0.0799$ and the distance modulus $\mu=37.77$ were obtained from the spectrum and the extinction was deduced from \cite{Schlafly2011} and using the extinction curves of \cite{Cardelli1989}. All these parameters are summarized in Table~\ref{table:param}.

Previous PTF observations were obtained in the years prior to the SN explosion, but no previous detection of any precursor outbearst exist. The most recent non-detection was on 2012 June 25. We present a derivation of the explosion epoch in \S~\ref{sec:bolo}. 

\begin{deluxetable}{lr}
\tablecolumns{2}
\tablewidth{0pt}
\tablehead{\colhead{Parameter}&\colhead{Value}}
\startdata
right ascension $\alpha$ ($J2000$)& $238.721000$\,deg \\
declination $\delta$ ($J2000$)& $3.535421$\,deg\\ 
redshift $z$&$z = 0.0799$\\
distance modulus $\mu$&$37.77$\,mag\\ 
galactic extinction $E_{B-V}$&$0.13$\,mag\\  
\enddata
\tablecomments{Summary of PTF\,12glz observational parameters.} 
\label{table:param}
\end{deluxetable}

\subsection{Photometry}

PTF\,12glz was observed in multiple bands for almost three years after discovery. 
The SN was monitored during a rising phase ($t<36$ days) and a decay phase ($t>243$ days) but not around peak luminosity. All the host-subtracted light curves are shown in Figure~\ref{fig:lc}. The photometry is reported in electronic Table~\ref{table:photo} and is available via WISeREP\footnote{https://wiserep.weizmann.ac.il}. 

\textit{GALEX} observations of the PTF\,12glz field started on 2012 May 26 and 15 observations were obtained with a cadence of $\sim3$ days. The \textit{GALEX NUV} camera 
was operating in scanning mode and observed strips of sky in a drift-scan mode with an effective average integration time of $80\,\rm{s}$, to a $NUV$ limiting magnitude of 20.6 mag [AB]. The \textit{GALEX} data reduction was done using tools\footnote{MATLAB Astronomy \& Astrophysics Toolbox, https://webhome.weizmann.ac.il/home/eofek/matlab/} by \cite{Ofek_matlab}. 

The P48 telescope was used with a $12$K$\times 12$K CCD mosaic camera \citep{Rahmer2008} and a Mould $R$-band filter. Data were obtained with a cadence of $\sim2$ days, to a limiting magnitude of $\rm{R}\approx21$ mag [AB]. For the data reduction of the P48 data, we used a pipeline developed by Mark Sullivan \citep{Sullivan2006,Firth2015}. 

The robotic $1.52$\,m telescope at Palomar (P60; \citealt{Cenko2016}) was used with a $2048\times2048$-pixel CCD camera and $g'$, $r'$, $i'$ SDSS filters. Data reduction of the P60 data was performed using the FPipe pipeline \citep{Fremling2016}.
We calibrated the P60 data in the following way. The $r$-band light curve was scaled so that its average value during the time window covered by both telescopes matches the average value of the P48 $R$-band photometric data. The $g$-band and $i$-band data were scaled to match the synthetic photometry of the calibrated spectroscopic data (\S~\ref{sec:obs-spectroscopy}). The synthetic photometry used for the calibration and for other purposes in this paper was computed with the PyPhot\footnote{http://mfouesneau.github.io/docs/pyphot/} pipeline (Fouesneau, in preparation).

Although the photometric data available for PTF\,12glz do not cover the peak, the data during the rise and decay allow to place an upper limit on the absolute magnitude at peak: with  $M_r\lesssim-20$, PTF\,12glz is at the bright-end of the observed SNe IIn, together with e.g., SN 2006gy \citep{Ofek2007,Smith2007}, SNe 2008fq \citep{Thrasher2008,Taddia2013} or SN\,2003ma \citep{Rest2009,Rest2011}. In particular, it is brighter than all SNe in the sample by \cite{Kiewe2012}, which was designed to be unbiased.

\begin{deluxetable}{llll}
\tablecolumns{4}
\tablewidth{0pt}
\tablecaption{}
\tablehead{\colhead{Epoch (days)}&\colhead{Counts (arb.)}&\colhead{Mag (magAB)}&\colhead{Instrument}}
\startdata
$11.24$  &$0.75\pm0.13$&$20.38\pm0.28$ & \textit{GALEX}nUV\\
$9.71$ &$362.30\pm86.30$&$20.60\pm0.26$& P48/R\\
$185.48$ &-&$19.01\pm0.06$ & P60/g'\\
$185.48$ &-&$18.85\pm0.04$ & P60/r'\\
$185.48$ &-&$18.40\pm0.05$ & P60/i'
\enddata
\tablecomments{{\bf Photometry.} This table is available in its entirety in machine-readable format in
the online journal. A portion is shown here for guidance regarding its form and content. Time is shown relative to the estimated epoch at which the extrapolated light curve (based on Equation~\ref{eq:bolometric luminosity}) is crossing zero: $t_{0}=2456097.58$ (2012 June 19), as derived in \S~\ref{sec:bolo}. To compute the apparent magnitudes from the counts, the zero-point for the $nUV$ data is $ZP_{nUV}=20.08$ and the zero-point for the P48 data is $ZP_{P48}=27.00$.} 
\label{table:photo}
\end{deluxetable}

\begin{figure*}
\begin{center}
\includegraphics[scale=.80]{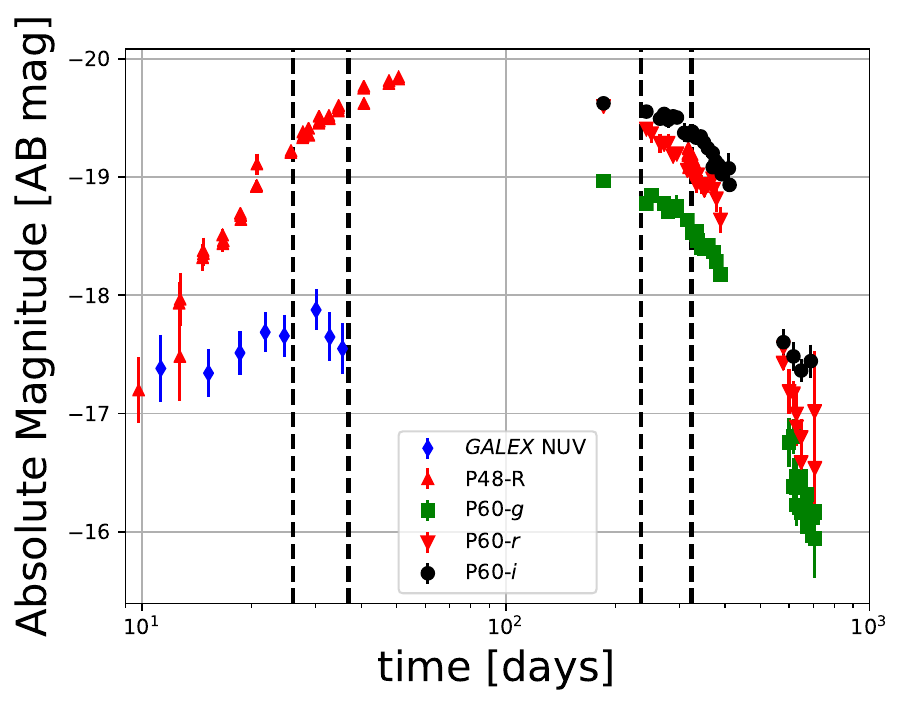}
\caption{The light curve of PTF\,12glz. Time is shown relative to the estimated epoch at which the extrapolated light curve (Equation~\ref{eq:bolometric luminosity}) is crossing zero: $t_{0}=2456097.58$ (2012 June 19), as derived in \S~\ref{sec:bolo}. Black dashed lines indicate dates at which spectroscopic data exist.} 
\label{fig:lc}
\end{center}

\end{figure*}

\subsection{Spectroscopy}\label{sec:obs-spectroscopy}

Four optical spectra of PTF\,12glz were obtained using the telescopes and spectrographs listed in Table~\ref{table:obs}.
The two first spectra were taken during the light curve rise, and the two last ones during the decay, at the dates shown in  Table~\ref{table:obs}. The spectra were used to determine the redshift $z = 0.0799$ from the narrow host lines (H${\rm \alpha}$ and [OIII]). All the observations were corrected for a galactic extinction of $E_{B-V}=0.13$\,mag, deduced from \cite{Schlafly2011} and using \cite{Cardelli1989} extinction curves, with the parameter $R\equiv A(V)/E_{B-V}$ (i.e. the ratio of total to selective extinction at V) set to the value $R=3.1$. In \S~\ref{sec:peculiar_r}, we show that our qualitative results are not affected when varying $R$, e.g. within the interval given in \cite{Fitzpatrick1999}.  

The spectroscopic observations were 
calibrated in the following way: the first two and the last spectra, for which we have contemporaneous P48 $R$-band data, were scaled so that their synthetic photometry matches the P48 $R$-band value. The third spectrum was scaled in the same way using the overlapping P60 $r$-band data instead. 

\begin{figure*}
\begin{center}
\includegraphics[scale=.55]{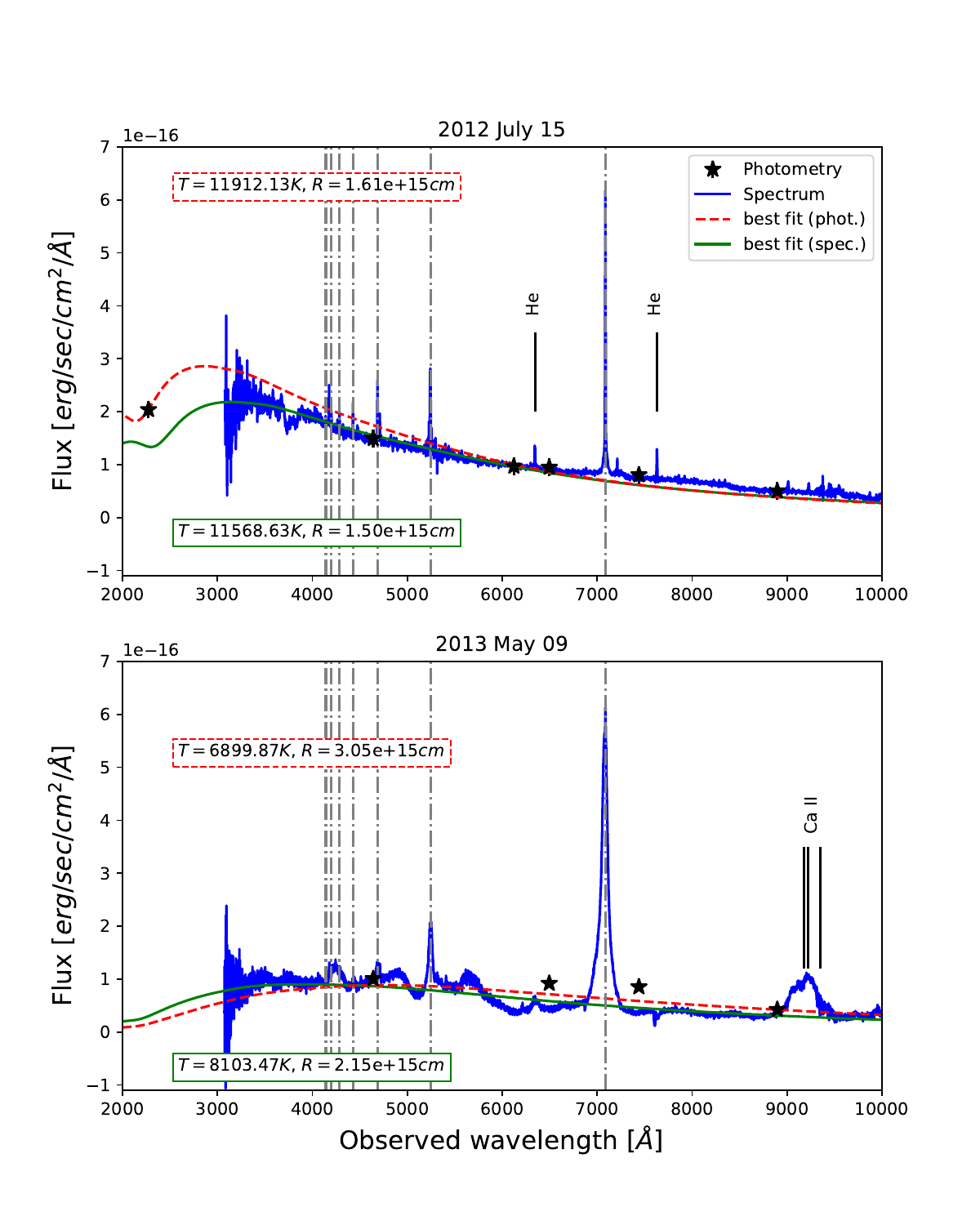}
\caption{The earliest (top) and latest (bottom) observed spectra of PTF\,12glz. Both spectra were calibrated to the $R$-band photometric measurement. Dashed lines indicate the redshifted emission lines for the Balmer series. Black stars show the combination of the observed ($NUV$ and P48 $R$-band) and synthetic ($g$, $r$ and $i$ sdss bands) photometry. The dashed red line shows the blackbody curves that best fits the photometric data (and the best fit values are shown in the box with dashed contours), while the green continuous line shows the blackbody curve that bests fit to the spectroscopic data (and the best fit values are shown in the box with dashed contours). (A color version of this figure is available in the online journal).}
\label{fig:spectra}
\end{center}
\end{figure*}

The first and last spectra are shown in Figure~\ref{fig:spectra} (the first two spectra are very similar, as well as the last two spectra) and all spectra are available from the Weizmann Interactive Supernova data REPository\footnote{https://wiserep.weizmann.ac.il} (WISeREP, \citealt{Yaron2012}).


\begin{deluxetable}{llll}
\tablecolumns{4}
\tablewidth{0pt}
\tablehead{\colhead{Date}&\colhead{Phase}&\colhead{Facility}&\colhead{Reference}}
\startdata
2012 July 15&$+26.2$ days&P200 &\cite{Oke1982}\\
2012 July 26&$+36.9$ days&P200 &-\\
2013 February 9&$+235.1$ days&LRIS&\cite{Oke1994}\\
2013 May 9&$+324.0$ days&LRIS&-\\
\enddata
\tablecomments{Spectroscopic observations of PTF\,12glz. The phase is given from the explosion epoch derived in \S~\ref{sec:bolo}. The double-beam spectrometer \citep{Oke1982} mounted on the $200$ " Hale telescope at Palomar was used with a 1'' slit, the $5500\rm \AA$ dichroic and the $316/7150$ grating positionned at a grating angle of $24\,\rm deg\,38.2\, \rm min$. The Low-Resolution Imaging Spectrometer (LRIS) \citep{Oke1994} spectrometer mounted on the $10$\,m Keck I telescope was used with a 1'' slit, the $5600\rm \AA$ dichroic and the $400/3400$ grism on the blue side.} 
\label{table:obs}
\end{deluxetable}

\section{Analysis}

\subsection{Spectroscopy}\label{sec:analysis_spectroscopy}


The two early spectra, obtained during the rise of the light curve, are characteristic of interacting SNe: a blue continuum with strong and narrow Balmer lines, as well as weak He I ($5876$\AA, $7065$\AA) narrow lines. At short wavelengths, the spectra show absorption from iron, as seen, for example, in SN\,2010jl\footnote{https://wiserep.weizmann.ac.il}.

We fitted a blackbody spectrum to the six-point spectral energy distribution (corrected for redshift and extinction) obtained by combining (1) the observed photometry in the $NUV$ and P48 $R$-band (2) the synthetic photometry of the spectra in the $g$, $i$ and $r$ SDSS bands. 
The best-fit temperatures and radii are shown in Table~\ref{table:sp-analysis} and Figure~\ref{fig:evo_param_TR} (as stars). In Figure~\ref{fig:spectra}, we show the synthetic and observed photometry derived for the earliest spectrum, on which is superimposed the calibrated spectrum and the best blackbody fit. 

\begin{figure}
\includegraphics[scale=.55]{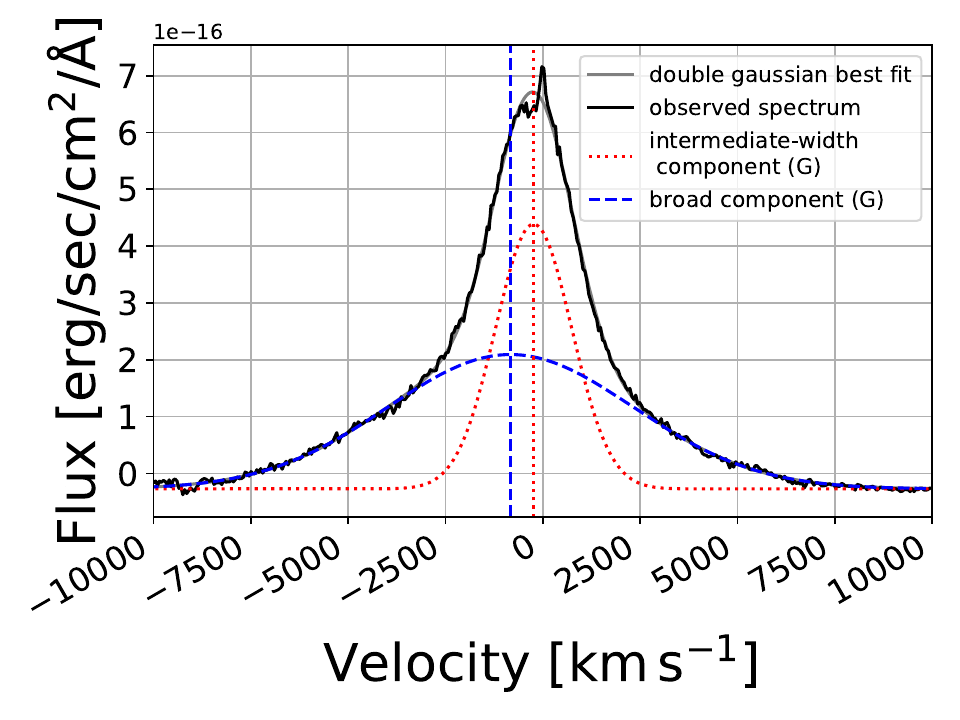}
\caption{The H${\rm \alpha}$ line profile on 2013 May 09 (latest spectrum). The bold black line shows the observed spectrum. The red dotted line and blue dashed line are the  intermediate-width and broad Gaussian (G) components whose linear combination (in grey) best fits the data. The vertical lines indicate the center of each component: the narrow component is blueshifted by $\approx-230$\,km\,s$^{-1}$ and the intermediate-width component is blueshifted by $\approx-830$\,km\,s$^{-1}$.} 
\label{fig:broad_lines}
\end{figure}

Both early spectra  show strong and narrow Balmer lines, which for SNe IIn are interpreted as coming from the  slow, unshocked, photoionized CSM. Their broad Lorentzian wings may be the signature of electron scattering, as the H${\rm \alpha}$ photons diffuse ahead of the shock through the dense CSM (e.g., \citealt{Chugai2001}). After subtracting the best-fit continuum from the spectra, we fitted the narrow H${\rm \alpha}$ lines. We tried several linear combinations of Gaussian and Lorentzian functions: the best fit is a superposition of a narrow Gaussian component with $\rm FWHM\approx100 - 200$\,km\,s$^{-1}$ (i.e., unresolved), which we interpret as tracing the slow unshocked CSM and an intermediate Lorentzian component with $\rm FWHM\approx500 - 700$\,km\,s$^{-1}$ (if some of the line-broadening comes from electron scattering, this is an upper limit of the CSM speed). The derived speeds and offsets are shown in Table~\ref{table:sp-analysis}. Figure~\ref{fig:broad_lines} shows the line and the best fit for the latest spectrum. 

No signature of expanding material is visible in the spectrum at this stage, which is consistent with a thick CSM obstructing the view of the SN ejecta at early times. We will show in \S~\ref{sec:peculiar_r} how the photometric data are inconsistent with this picture.

The fits mentioned above, as well as all those mentioned in the rest of this paper, were performed using the {\tt emcee} algorithm \citep{Foreman2013} to sample from the posterior probability distribution. 
We then used the ten combinations with the lowest $\chi^2$ from the Monte Carlo Markov chain as initial conditions for an optimization algorithm to compute the best fit value (the maximum {\it a posteriori} value or "m.a.p.", in the terminology of \citealt{Hogg2010}). Note that we used MCMC to compute the posterior distribution and deduce the error bars on the one hand and on the other hand, used a distinct optimization algorithm to solve for the set of parameters maximizing the posterior, i.e. find the m.a.p. (which coincides with the maximum likelihood value and the minimum $\chi^2$ value since the priors are uninformative). The reason we took this precaution - which turned out to be un-necessary - is that in the case of non-gaussian or non-symmetric marginalized posterior distribution, the m.a.p. may not fall necessary close to the median of the posterior distribution. In such cases, computing the position of the m.a.p. can be problematic and challenging, all the more when the $\chi^2$ is noisy and full of local minima. The main challenge is then to choose an initial combination of parameters values to give a minimization algorithm (most minimization algorithms, require to be given an initial combination of parameters from which to start their search for the minimum). The $10$ sets of parameters from the chain giving the lowest $\chi^2$ is a reasonable yet somewhat arbitrary choice of such initial condition. In the current case, since the posterior distribution is reasonably close to a Gaussian and well-behaved, the m.a.p. coincides with the median of the posterior distribution.

When errors are noted, they correspond to the $1\sigma$ limits of the marginalized posterior distributions.

\begin{figure}
\includegraphics[scale=.55]{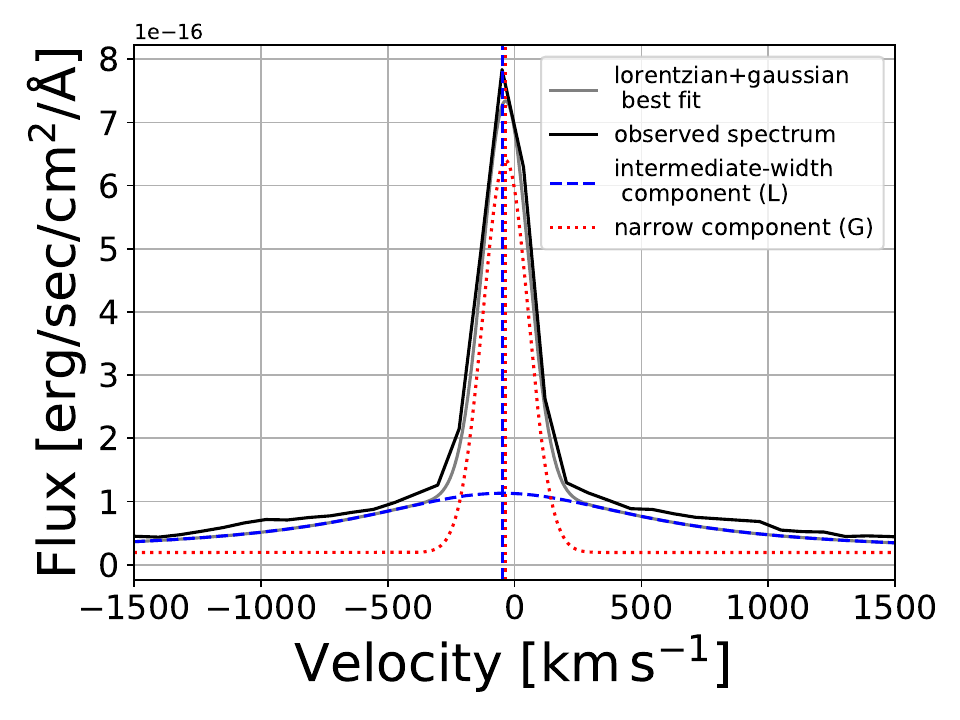}
\caption{The H${\rm \alpha}$ line profile on 2012 July 15 (earliest spectrum). The bold black line shows the observed spectrum. The red dotted line and blue dashed line are the narrow Gaussian (G) and intermediate-width Lorentzian (L) whose linear combination (in grey continuous line) best fits the data. The origin of the x-axis indicates the location of the H${\rm \alpha}$ line at the host galaxy, at redshift $z=0.0799$. The vertical lines indicate the center of each component: the narrow component is blueshifted by $\sim-30$\,km\,s$^{-1}$ and the intermediate-width component is blueshifted by $\sim-50$\,km\,s$^{-1}$.} 
\label{fig:narrow_lines}
\end{figure}


The interpretation of the late-time spectra of Type IIn SNe should be made with the complexity caused by CSM interaction in mind. In ``normal'' type II SNe, the ejecta that become optically thin at late times are heated from the inside by two sources of energy: (1) remaining thermal deposition from the original heat of the explosion and (2) radioactivity. The late, or ``nebular'' spectrum shows no clear continuum, and its emission lines reflect the expansion of the ejecta in which they formed. In SNe IIn, the CSM interaction may continue to dominate the spectrum at late times, e.g., because the ejecta is heated by the shock wave propagating backward from the CSM into its outermost layers \citep{Chevalier2003}. 

The two late-time spectra of PTF\,12glz look similar (the latest spectrum is shown in Figure~\ref{fig:spectra}): they show a weak continuum in the red, a pseudo-continuum presumably formed by the superposition of narrow [Fe II] emission lines in the blue (e.g., \citealt{Kiewe2012}) and several broad emission lines. The temperatures and radii derived by fitting a blackbody curve to the observed spectrum are listed in Table~\ref{table:sp-analysis} and compared to the parameters derived from photometry in Figure~\ref{fig:evo_param_TR}. 
Both spectra show broad Balmer emission lines. The derived speeds and redshifts are presented in Table~\ref{table:sp-analysis} and Figure~\ref{fig:broad_lines} shows the line and the best fit for the latest spectrum. The H${\rm \alpha}$ lines are best fit by a linear combination of an intermediate-width Gaussian component with $FWHM\approx2000$\,km\,s$^{-1}$ that is offset to the blue by $\approx250$\,km\,s$^{-1}$ relative to the galaxy rest-frame and a broad Gaussian component with $FWHM\approx8000$\,km\,s$^{-1}$ that is offset to the blue by $\approx1000$\,km\,s$^{-1}$ relative to the galaxy rest frame. The intermediate-width component could come from the shocked gas in a cold dense shell forming at the contact discontinuity between the decelerated ejecta and the shocked CSM, which is reheated by X-rays and $UV$ radiation from the shock. The broad component velocities are too high to originate from the CSM: they are rather characteristic ejecta velocity values. We deduce from these velocities that the ejecta has - at least partially - emerged through the optically thick layers of the CSM. 



Both late spectra also show emission from the the Ca II IR triplet. This is commonly seen in late time-interacting SNe, e.g., in the type IIn SN\,2005ip \citep{Boles2005,Stritzinger2012}, the type Ic SN\,2007dio \citep{Kuncarayakti2018} and the type Ia-csm PTF\,11kx \citep{Dilday2012,Silverman2013,Graham2017}, as well as in other types of SNe.



\begin{deluxetable*}{lllllllll}
\tablecolumns{9}
\tablewidth{0pt}
\tablecaption{}
\tablehead{Date &\multicolumn{2}{c}{Continuum} & \multicolumn{2}{c}{H${\rm \alpha}$ narrow comp.} &\multicolumn{2}{c}{H${\rm \alpha}$ intermediate comp.}&\multicolumn{2}{c}{H${\rm \alpha}$ broad comp.}}
\startdata
 &$T_{BB}$ &$R_{BB}$&$\Delta v$&$FWHM$&$\Delta v$&$FWHM$&$\Delta v$&$FWHM$ \\ 
 &[K]&[$10^{15}$cm]&[km\,s$^{-1}$]&[km\,s$^{-1}$]&[km\,s$^{-1}$]&[km\,s$^{-1}$]&[km\,s$^{-1}$]&[km\,s$^{-1}$]\\  \hline \\
2012 July 15&$11570^{+350}_{-400}$&$1.50^{+0.11}_{-0.07}$&$-35$&$110$&$-50$&$680$&-&- \\
2012 July 26&$8400^{+2780}_{-440}$&$2.91_{-0.92}^{+0.36}$&$-80 $&$240$&$-100$&$530$&-&-\\
2013 February 09&$8200^{+340}_{-350}$&$2.56_{-0.20}^{+0.24}$&-&-&$-270$&$1990$&$-1070$&$8580$ \\
2013 May 09&$8100^{+420}_{-390}$&$2.15_{-0.22}^{+0.26}$&-&-&$-240$&$2310$&$-830$&$7470$ \\
\enddata
\tablecomments{The table shows the best fit values of the continuum and H${\rm \alpha}$ lines in the four spectra of PTF\,12glz. The early spectra are best fit with a linear combination of a narrow Gaussian component (left column) and an intermediate-width Lorentzian component (central column). The late spectra are best fit with a linear combination of an intermediate-width Gaussian component (central column) and a broad Gaussian component (right column). $\Delta z$ is the shift of the centre of each component compared to the galaxy rest frame (the negative values mean that the component is blueshifted).} 
\label{table:sp-analysis}
\end{deluxetable*}

\subsection{Host galaxy}\label{sec:host}


We retrieved science-ready imaging data from the several surveys, summarized in Table \ref{tab:host_phot}: the Galaxy Evolution Explorer (\textit{GALEX} ) general release 7 (GR7) \citep{Martin2005a},
the Panoramic Survey Telescope and Rapid Response System (Pan-STARRS; PS1) data release 1 (DR1) \citep{Flewelling2016a} and
the Sloan Digital sky survey data release 9 (DR9) (SDSS; \citealt{Ahn2012a}). We used the software package {\tt LAMBDAR} (Lambda Adaptive Multi-Band Deblending Algorithm in R; \citealt{Wright2016a}) that is based on \citet{Bourne2012a} to perform multi-band matched aperture photometry (i.e., taking into account different pixel scales and point-spread functions). The absolute flux calibration was done against instrument-specific zero points (For details on the photometry see Schulze et al. in prep.).

\begin{deluxetable}{llll}
\tablecolumns{4}
\tablewidth{0pt}
\tablecaption{}
\tablehead{Instrument & Filter    & Wavelength    & Brightness }
\startdata               
 &           & (\AA)         & (mag$_{\rm AB}$)  \\

     \textit{GALEX}      & FUV       & 1542.3        & $21.55\pm0.46$    \\
    \textit{GALEX}      & NUV       & 2274.4        & $21.23\pm0.51$    \\
    SDSS        & $u'$      & 3594.9        & $19.88\pm0.17$    \\
    SDSS        & $g'$      & 4640.4        & $18.59\pm0.07$    \\
    SDSS        & $r'$      & 6122.3        & $17.96\pm0.08$    \\
    PS1         & $i'$      & 7439.5        & $17.75\pm0.07$    \\
    SDSS        & $z'$      & 8961.5        & $17.59\pm0.10$    \\
    PS1         & $Y$       & 9603.1        & $17.23\pm0.19$    \\
\enddata
\tablecomments{{\bf Photometry of the host galaxy.} The photometry is not corrected for Galactic reddening. The effective wavelengths were taken from the Spanish Virtual Observatory\footnote{\href{https://svo.cab.inta-csic.es/}{https://svo.cab.inta-csic.es/}}.} 
\label{tab:host_phot}
\end{deluxetable}

We modelled the spectral energy distribution (SED) of the host galaxy with the software packages {\tt Le Phare} on the web portal {\tt GAZPAR}\footnote{\href{https://gazpar.lam.fr}{https://gazpar.lam.fr}} (GAlaxy photometric redshifts (Z) \& physical PARameters). We modelled the data using a grid of templates based on \citet{Bruzual2003a} stellar population-synthesis models with the Chabrier initial-mass-function \citep{Chabrier2003a}, a star-formation history that is approximated by a declining exponential function, and a \citet{Calzetti2000a} dust attenuation curve.

\begin{figure}[h]
    \centering
    \includegraphics[width=1\columnwidth]{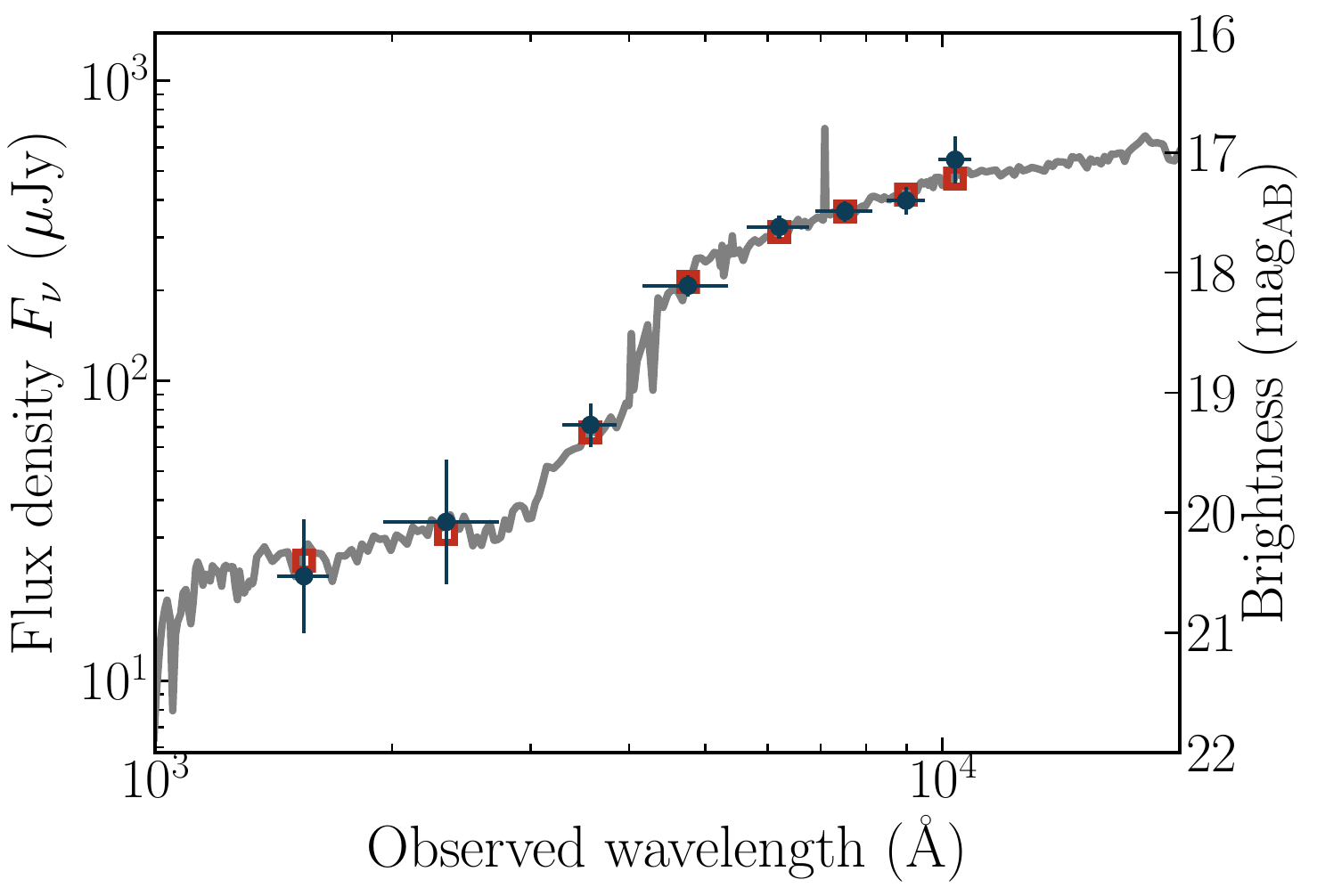}
    \caption{Spectral energy distributions of the host of PTF12glz from 1000 to 14000 \AA. The observed magnitudes are displayed by the circles. The solid line displays the best-fit model of the SED with {\tt Le Phare}. The squares are the model predicted magnitudes.
    }
    \label{fig:host_sed}
    \end{figure}

Figure \ref{fig:host_sed} shows the measured galaxy photometry and the best fit. The SED can be adequately described by a template with a stellar mass of $\log M/M_\odot = 9.0^{+0.7}_{-0.2}$, a star-formation rate of $1.9^{+4.1}_{-0.7}~M_\odot\,{\rm yr}^{-1}$ and negligible attenuation $\left[E\left(B-V\right) = 0\right]$ ($\chi^2$/number of filters = 1.3/8). Using the parametrisation of the mass-metallicity relation by \citet{Andrews2013a}, we estimate the galaxy metallicity to be $\sim0.4$--0.5 solar.

\subsection{The peculiar evolution of the blackbody radius}\label{sec:peculiar_r}

\begin{figure}\label{fig:evo}
\begin{center}
\includegraphics[scale=.43]{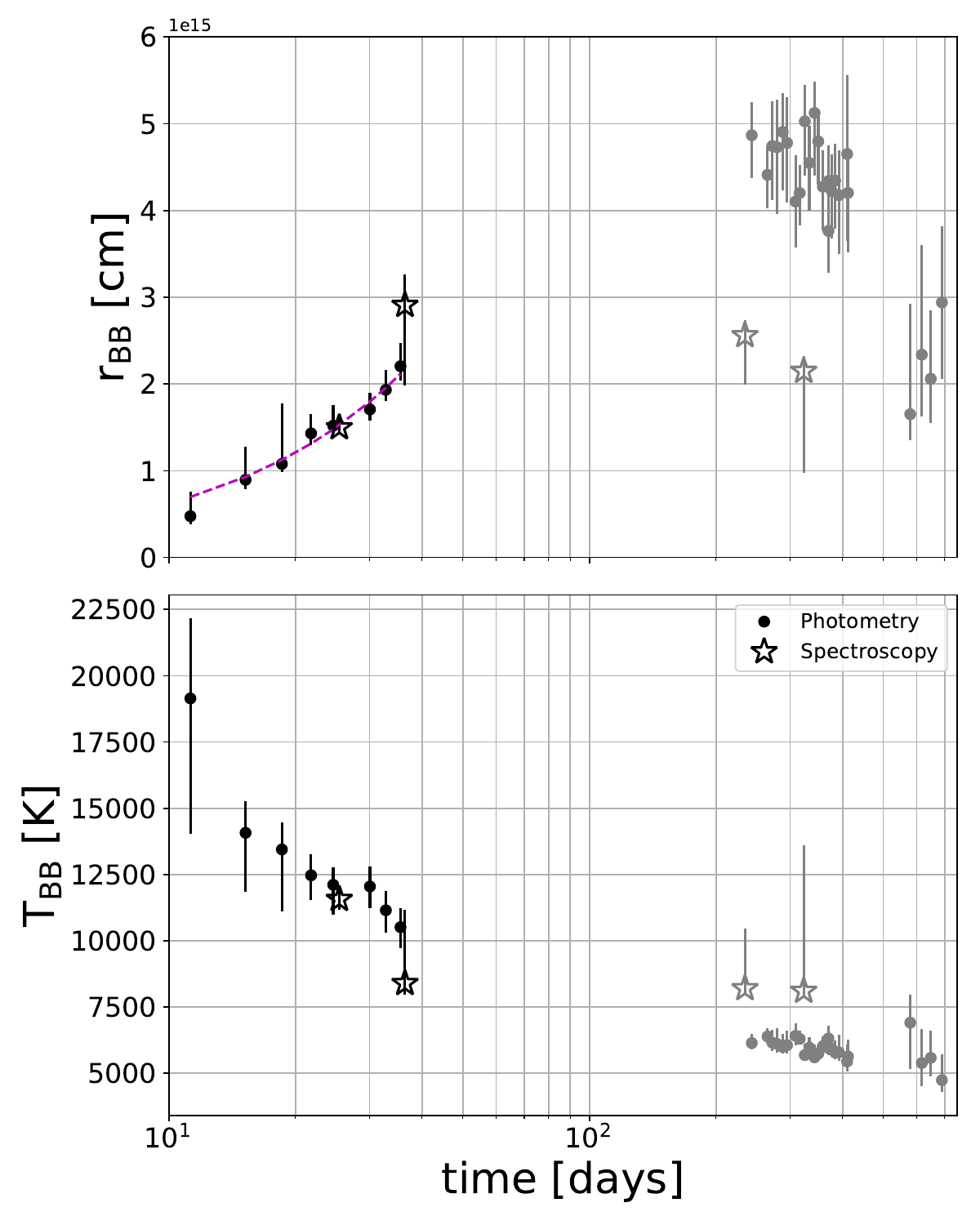}
\caption{The evolution in time of: (1) the radius (top panel), (2) the temperature (lower panel) of a blackbody with the same radiation as PTF\,12glz. The points were obtained by fitting a blackbody spectrum to the observed photometry, after interpolating the various data sets to obtain data coverage of coinciding epochs. The errors were obtained with MCMC simulations. The star symbols indicate the values derived by fitting a blackbody to the spectroscopic data. The dashed line in the top panel shows the best linear fit to the rising radius phase: a linear function with a slope of $\approx7000$\,km\,s$^{-1}$. At late times, the blackbody model for the spectral energy distribution may not be valid anymore (see, e.g., right panel in Figure~\ref{fig:spectra}): these points are shown in grey to emphasise that they are less reliable and should be taken cautiously.} 
\label{fig:evo_param_TR}
\end{center}
\end{figure}

Taking advantage of the multiple-band photometry coverage, we derived the temperature and radius of the blackbody that best fits the photometric data at each epoch (after correcting for redshift and extinction, interpolating the various data sets to obtain data coverage of coinciding epochs, and deriving the errors at the interpolated points with Monte Carlo Markov chain (MCMC) simulations). The derived best-fit temperatures $T_{BB}$ and radii $r_{BB}$ are shown in Figure ~\ref{fig:evo_param_TR}. For the temperature MCMC fit, we adopted a broad, uninformative (flat) prior $T \in [5000, 25000]$. The edge values of the prior were chosen to contain the range of temperature observed e.g., in the SNe IIn sample by \cite{Taddia2013}. The best fit temperatures $T_{BB}$ should be seen as a lower limit on the temperature, since the spectra of SNe typically show a deficit of flux due to line-blanketing by metal lines.

In most epochs, this method implies fitting a blackbody spectrum to a two-point spectral energy distribution. 
Comparison with results derived from more constraining data detailed in \S~\ref{sec:analysis_spectroscopy} - either spectroscopy (shown in red) or a combination of observed and synthetic photometry (shown in green) - suggests that this method leads to a slight overestimation of the radius and underestimation of the temperature.

The temperature $T_{BB}$ drops from $\approx15000$\,K to $\approx8000$\,K during the rise and is stable at $\approx6000$\,K during the decay phase. 
The decrease of $T_{BB}$ at early time is well fitted by a power law $t^n$ with index $n=-0.6$ and is consistent with the temperature evolution observed in the sample by \cite{Taddia2013}. However, PTF\,12glz is relatively hot compared to the SNe IIn of this sample, where temperatures span between $11500$\,K and $5500$\,K, and compared to other well studied SNe IIn (e.g., 2006gy, \citealt{Ofek2007, Smith2007}; SN 2005ip, \citealt{Smith2009b}; SN 2010jl, \citealt{Ofek2014}).

The derived radius grows by an order of magnitude, from $r_{BB}\approx6\times10^{14}$ cm to $r_{BB}\approx3\times10^{15}$ cm during the rising phase and is $\approx4\times10^{15}$ cm during the light curve decay phase. Such a rise is puzzling within a picture where the optically thick CSM is supposed to mask the shock and the expanding material. To our knowledge, the measured blackbody radius of all SNe IIn observed to date either stalls after a slight increase (e.g., 2005kj, 2006bo, 2008fq, 2006qq, \citealt{Taddia2013}; 2006tf, \citealt{Smith2008}), or stays relatively constant at early times (e.g., SN2010jl \citealt{Ofek2014}), or even supposedly shrinks (e.g., SN2005ip; SN2006jd, \citealt{Taddia2013}). Whereas a constant radius is consistent with the continuum photosphere being located in the unshocked optically thick CSM, the possible presence of clumps in the CSM that may expose underlying layers, has been invoked by \cite{Smith2008} to interpret observations of a stalling or shrinking radius.

In our case, the velocity at which $r_{BB}$ grows, reaches $\approx7000$\,km\,s$^{-1}$, a value too large to trace the unshocked layers of the CSM or even the reverse shock. One may naively think that if the CSM was optically thin, then the observed radius would grow. However, in this case, it would be hard to convert efficiently the hard X-ray produced by the colisionless shock into optical radiation. In \S~\ref{sec:discussion} we propose one possible geometrical solution to this puzzle.

To investigate the effect of variations in the extinction and mimic the impact of errors in $E_{B-V}$ on this results, we re-ran the analysis for the two extreme values of the parameter $R\equiv A(V)/E_{B-V}$ mentioned in \cite{Fitzpatrick1999}: 2.2 and 5.8. We observed an offset in the radius and temperature but our qualitative result - a growth in the radius at a velocity $\gtrapprox7000$\,km\,s$^{-1}$ and a decrease in the temperature - is maintained.

\subsection{Bolometric light curve}\label{sec:bolo}
\begin{figure}
\begin{center}
\includegraphics[scale=.55]{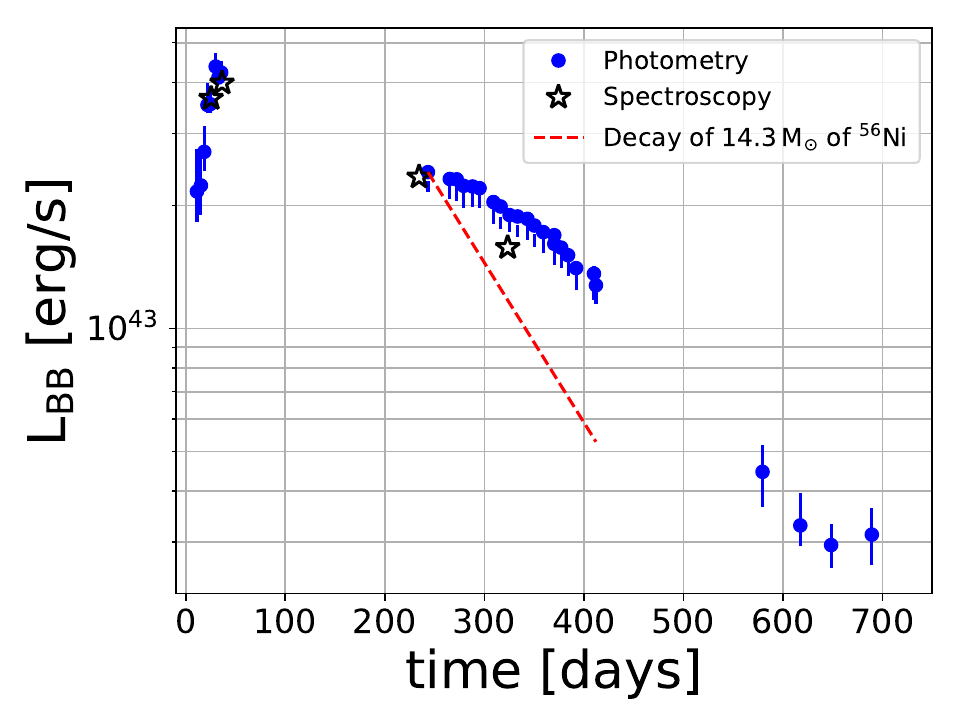}
\caption{The evolution in time of the bolometric luminosity of a blackbody with the same radiation as PTF\,12glz. The star symbols indicate the fits to the spectroscopic data. The dashed line shows the variation in luminosity caused by $14.3\;\rm{M_{\odot}}$ of ${}^{56}$Ni radioactively decaying to ${}^{56}$Co and then to ${}^{56}$Fe. $L_{BB}$ decays $\approx2$ times slower, which suggests that radioactive decay is not sufficient to explain the decay of the light curve and that interaction or other sources of radiation play a role.}
\label{fig:evo_param_L}
\end{center}
\end{figure}

Based on the measurement of $r_{BB}$ and $T_{BB}$, we were able to derive the luminosity $L_{BB}=4\pi R^2\sigma T^4 $ of the blackbody fits, shown in Figure~\ref{fig:evo_param_L}. Since PTF\,12glz was not observed at peak luminosity, we can only derive a lower limit on the peak luminosity $L_{BB,peak}>4\times10^{43}\,\rm erg\,s^{-1}$. This makes PTF\,12glz more luminous than SN\,2008fq - the brightest SN of the \cite{Taddia2013} sample - and brighter than all but one SNe of the sample by \cite{Ofek2014b}. This suggests that PTF\,12glz is at the bright end of the SNe IIn luminosity range. 

We fitted the light curve during the rise time with a function of the form
\begin{equation}\label{eq:bolometric luminosity}
L=L_{\rm{max}}\{ 1-\exp{[(t_{0}-t)/t_c]} \}\;,
\end{equation}
(where $t_{0}$ is the time of zero flux, $L_{\rm{max}}$ is the maximum bolometric luminosity, and $t_{c}$ is the characteristic rise time of the bolometric light curve). This allowed us to estimate the epoch at which the extrapolated light curve is crossing zero, which is used throughout this paper as the reference time $t_{0}\,\rm{(MJD)}=56097.58$.

As shown in Figure~\ref{fig:evo_param_L}, $L_{BB}$ decreases about twice as slowly as the 0.98\,mag/100\,d decline rate characteristic of the radioactive decay of ${}^{56}$Ni to ${}^{56}$Co and then to ${}^{56}$Fe. If this decline was produced by the radioactive decay of ${}^{56}$Ni, a Nickel mass of at least $14\,\rm{M}_{\odot}$ would be required to reach the bolometric luminosity of PTF\,12glz (see Figure~\ref{fig:evo_param_L}). The evolution of $L$ should be taken cautiously, since at late time, a blackbody model for the spectral energy distribution may not be valid anymore, and so the temperatures and radii used to calculate $L$ are less reliable.
Although the late spectra analyzed in \S~\ref{sec:analysis_spectroscopy} suggest that the ejecta may have emerged through the CSM at late times, the slow decline of $L_{BB}$ hints at the fact that interaction may still play an important role in the radiation budget (this may happen through radiation from the reverse shock, or through processed radiation through the edge of the wind, or if the ejecta has only partially emerged through the CSM and is still interacting with it in some places).

\subsection{Mass of the CSM}\label{sec:mass}

A very crude estimation of the swept-up CSM mass can be obtained by assuming that the ejecta mass is comparable to the mass of the CSM and that the bolometric luminosity $L$ is accounted for by the conversion of the kinetic energy of the ejecta into radiation:
\begin{equation}\label{eq:crude}
\int_{t}L(t)dt\lesssim\frac{1}{2}M_{\rm{CSM}}v_e^2\;,
\end{equation}
where $M_{\rm{CSM}}$ is the mass of the CSM and $v_e$ is the velocity of the ejecta.
To an order of magnitude, this estimate will not be very different than estimates based on more realistic treatment of the hydrodynamics (e.g., \citealt{Ofek2014}).
We use the width $FWHM\approx8000$\,km\,s$^{-1}$ of the broad H${\rm \alpha}$ component at late time (see section~\ref{sec:analysis_spectroscopy}) as an approximation of the ejecta velocity $v_e$.
From the measured bolometric luminosity shown in Figure~\ref{fig:evo_param_L}, $\int_{t}L(t)dt\approx10^{51}$ erg. Substituting this value in Equation~\ref{eq:crude} gives $M_{\rm{CSM}}\gtrsim 2M_{\odot}$.

In interacting SNe, the optically thick CSM masks the explosion, which may leave an ambiguity about the underlying explosion type. In particular, type Ia SNe and core-collapse SNe exploding inside a thick CSM would result in similar observational signatures. The spectra would look similar at early time (as long as the explosion is masked by the CSM) and at very late times, when Ni has decayed and there is no more energy to illuminate the ejecta and create absorption lines in the spectrum. Here, the high value of the bolometric luminosity excludes the possibility that PTF\,12glz is a masked SN Ia.

Another order of magnitude estimate of the mass can be obtained by assuming a wind density profile, $\rho_{\rm{CSM}}(r)=K\,r^{-2}$ and using the photon diffusion timescale (e.g., \citealt{Ofek2010}),
\begin{equation}\label{eq:t_d}
t_{d}\sim\frac{\kappa K}{c}\;,
\end{equation}
where $\kappa$ is the CSM opacity.
We assume $t_d\sim t_c$, where $t_c$ is the characteristic rise time of the bolometric light curve. We estimate $t_c$ by fitting the rising part of the bolometric light curve with the exponential function defined in Equation~\ref{eq:bolometric luminosity} and assuming $\kappa\approx0.34$\,cm$^2$\,g$^{-1}$.
We obtain $t_{c}\sim20$ day
, and $K\approx1.5\times10^{17}\,\rm{g\,cm}^{-1}$, which corresponds to high values of the density $\rho_{\rm{CSM}}\gtrsim10^{-14}\,\rm{g\,cm}^{-3}$ (or a particle density of $n\sim10^{10}\,\rm{cm}^{-3}$, assuming a mean number of nucleons per particle $\langle\mu_p\rangle=0.6$) at the radii shown in Figure~\ref{fig:evo_param_TR}.

Measuring $K$ allows us to estimate the mass of the CSM swept up by the ejecta, $M_{\rm{CSM}}(r)\sim4\pi K r$ (assuming that the CSM extends on much higher scales than the stellar radius). Using the highest early-time radius $r_{BB}$, shown in Figure~\ref{fig:evo_param_TR}, as a lower limit of the maximum size of the shell of CSM surrounding the explosion, gives $M_{\rm{CSM}}\gtrsim3\,\rm{M}_{\odot}$, in good agreement with the estimates obtained with Equation~\ref{eq:crude}.

For a wind density profile, $K=\dot{M}/(4\pi v_{w})$, where $v_w$ is the velocity of the CSM. Therefore, measuring $K$ also gives us an estimation of the mass-loss rate. Using the width of the narrow H${\rm \alpha}$ component during rise time, $FWHM\approx200$\,km\,s$^{-1}$, as a proxi of $v_w$ (see Table~\ref{table:sp-analysis}), we obtain a large mass-loss rate $\dot{M}\sim0.6\; M_{\odot}$ yr$^{-1}$, higher than the mass-loss rates observed in \cite{Taddia2013} and \cite{Kiewe2012} and comparable to the mass-loss rate of e.g., iPTF13z \citep{Nyholm2017}. Combining these estimates suggests that the CSM mass was ejected on a timescale of $1$ to $10$ years prior to the SN explosion.

We wish to emphasize that we made several simplifying assumptions in this section. In particular (1) we assumed a spherical symmetry of the CSM, (2) we assumed a wind profile of the CSM and (3) we assumed that the kinetic energy of the ejecta converts efficiently into radiation. Therefore, the numbers above have to be considered as order of magnitude estimates.


%




\subsection{Dust formation?}

At late times, the broad component of the H${\rm \alpha}$ Balmer line is blueshifted by $\approx1000$\,km\,s$^{-1}$ relative to the galaxy rest frame. The intermediate component is also blueshifted, but by a velocity of $\approx250$\,km\,s$^{-1}$ which is consistent with a typical stellar velocity within the galaxy (see Table~\ref{table:sp-analysis} and Figure~\ref{fig:broad_lines}). Several explanations have been proposed to explain the blueshift of emission line profiles in interacting SN. One possible explanation is the formation of dust which increasingly blocks the receding parts of the ejecta (e.g., as proposed in the case of the Type IIn SN\,2010\,jl, \citealt{Smith2012,Gall2014}). In our case, the blueshift of the broad component does not grow with time (see Table~\ref{table:sp-analysis}), meaning that if there is dust, it may have formed before the epoch of the first nebular spectrum.  In our case, testing the wavelength dependency of the blueshift - a blueshift caused by dust would be stronger at bluer wavelengths - is tricky because the blended iron emission lines mask the structure of the $H\beta$ line profile. We tried to apply several filters on this area, to separate the possible broad Balmer components from Fe II blend structures, but the results remained inconclusive.

Other explanations have been proposed for the blueshift of emission lines. 
\cite{Fransson2014}, for example, attributed the blueshift of emission lines to radiative acceleration of the preshock gas by the SN radiation, whereas \cite{Smith2012} proposed a geometric explanation. 

\section{Radiative diffusion through a slab}\label{sec:diffusion}

The radiation from a SN exploding into a spherically symmetric CSM has been studied analytically, under simplifying assumptions (e.g., \citealt{Chevalier1982, Balberg2011,Ginzburg2014}) and numerically (e.g., \citealt{Falk1977, Moriya2014}). The case of an aspherical CSM has been explored to a lesser extent \citep{VanMarle2010, Mcdowell2018}. Exploring the expected effect of deviation from spherically symmetric CSM on the observables is all the more important since aspherical clouds of CSM around mass-ejecting stars seem to be common, e.g., in stars like $\eta$-Carinae \citep{Davidson1997} that have been proposed as SNe IIn progenitors (e.g., \citealt{Gal-Yam2007,Gal-Yam2009}). In this section, we attempt to determine whether a non-spherical geometry of the CSM around a SN can explain the growing radius and decreasing temperature observed in \S~\ref{sec:peculiar_r}. Solving for the exact shape of the CSM from a few observables in an ill-conditionned problem. Therefore, here our goal is merely to verify that a non-spherical geometry can explain the evolution of the observables. Given this goal, we consider a simple aspherical structure: a three-dimensional slab, infinite in two dimensions and perpendicular to the line of sight. 

\subsection{Model and assumptions}

\begin{figure}
\includegraphics[scale=.30, angle=270]{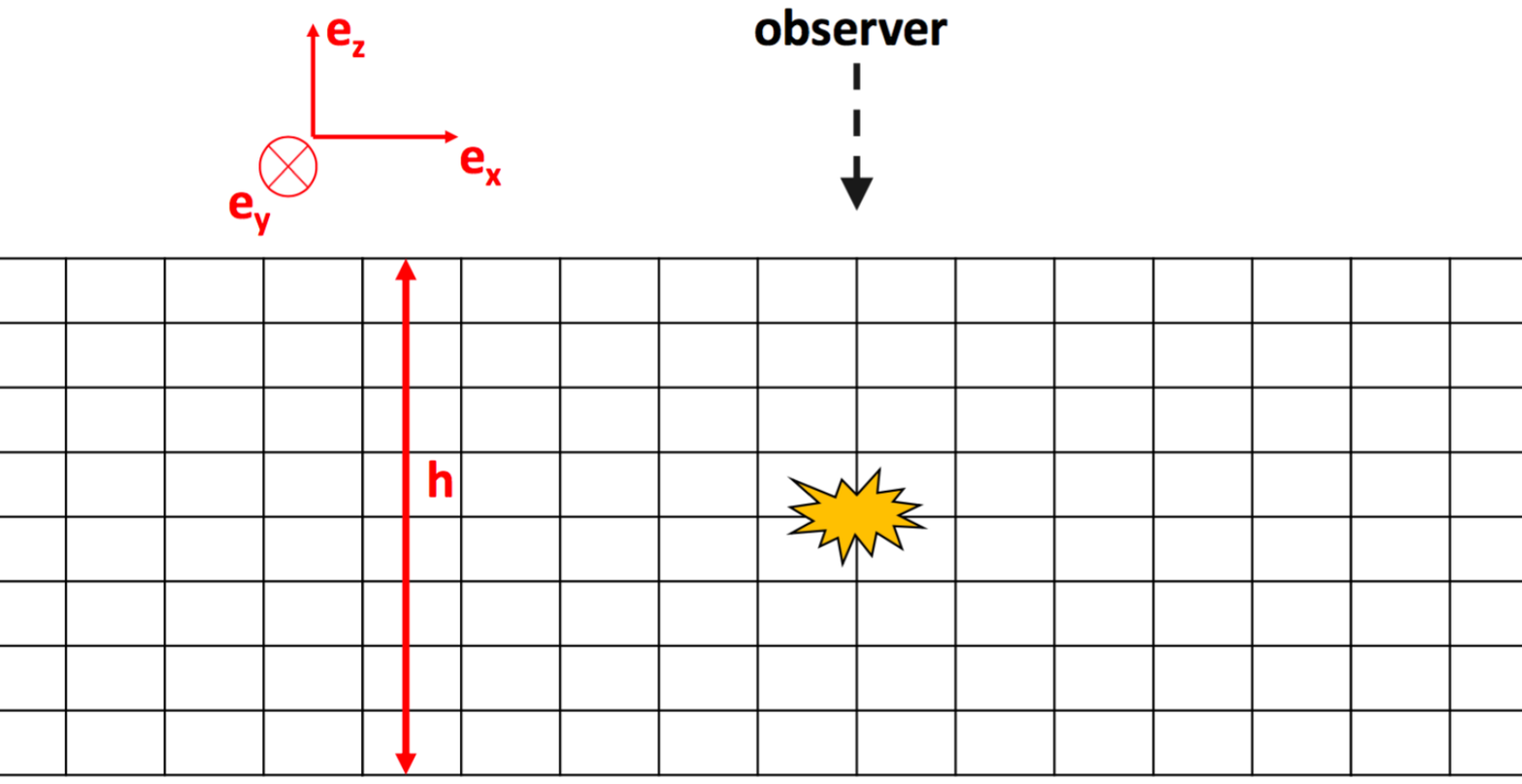}
\caption{Sketch of the grid used to simulate radiative diffusion through a slab. The dimensions of the slab along the $e_x$ and $e_y$ coordinates are much larger than $h$, the dimension of the slab parallel to the line of sight. The initial energy of the explosion is deposited at $z=0$, at a distance $h/2$ from the edge of the slab.} 
\label{fig:cartoon}
\end{figure}

We have written a three dimensional computer program in python,  \texttt{SLAB-Diffusion}, available on-line \citep[][Codebase: \url{https://github.com/maayane/SLAB-Diffusion}]{SLAB-Diffusion}, in order to calculate the propagation of photons through a slab and simulate the main observables. The simple geometry we consider is illustrated in a cartoon in Figure~\ref{fig:cartoon}. Following a similar approach as in \cite{Ginzburg2014}, we replaced the hydrodynamical description of the SN explosion by a stationary model of the shock breakout. This is equivalent to neglecting the expansion of the gas due to the explosion and modelling the interaction between the shock and the CSM as an instantaneous deposition of energy in the slab.

The assumption of an instantaneous release of energy is justified when $t_d/t_e\gg1$, where $t_e$ is the timescale over which the energy of the ejecta is converted into radiation by deceleration, and $t_d$ is the photon diffusion timescale (eq~\ref{eq:t_d}). The timescale $t_e$ is given by
\begin{equation}\label{eq:t_E}
t_e=\frac{R_c}{v}\;,
\end{equation}
where $R_c$ is the radius at which the accumulated CSM mass is comparable to the ejecta mass and $v$ is the ejecta velocity. Since $t_d/t_e=\tau v/c$, the condition $t_d/t_e\gg1$ is satisfied, and the instantaneous energy deposition approximation is valid, as long as $\tau\gg c/v\sim30$. This is indeed the case for the model parameters we infer for PTF\,12glz: a wind velocity $v=200\,\rm{km\,s^{-1}}$ (\S~\ref{sec:analysis_spectroscopy}) with a mass-loss rate $\dot{M}\sim0.6\,\rm{M_{\odot}}$ yr$^{-1}$ (\S~\ref{sec:mass}), yielding $R_c\sim10^{15}$\,cm and $\tau\sim200$.

We assumed that the problem can be treated accurately within the diffusion approximation (i.e., assuming that $\tau\approx1$ occurs close to the surface). Hence, the energy density $u$ within the slab is described by the diffusion equation:
\begin{equation}\label{eq:diffusion}
\frac{\partial u}{\partial t}=\nabla\cdot(D\vec{\nabla}u)\;,
\end{equation}
where $D= c/(3{\kappa \rho})$ is the diffusion coefficient. Here we explored three density profiles: a constant density profile, $\rho$, and two functions of the $\vec{e_z}$ coordinate: $\rho\propto |z|^{-1}$ and $\rho\propto z^{-2}$, where the origin of the $z$-axis is at the center of the slab. We leave the treatment of the angular dependency of $D$ to further extensions of this work.

At the boundaries, the energy escapes from the slab and the flux $\vec{f}$ is linked to the density of energy through
\begin{equation}
\lvert \lvert \vec{f}\rvert\rvert =\alpha c u\;,
\end{equation}
where $\alpha\approx1/4$ (the case $\alpha=1/4$ corresponds to isotropic radiation). 
We discretized Equation~\ref{eq:diffusion} in a cartesian, three-dimensional grid (illustrated in a cartoon in Figure~\ref{fig:cartoon}), using an explicit  forward Euler scheme with $\Delta t D/(\Delta d)^2<0.1$ where $\Delta d$ is defined for each coordinate and is the size of the mesh in each direction $\vec{e_x}$, $\vec{e_y}$ and $\vec{e_z}$. 
We assumed that $D$ does not depend on the wavelength of the photons (i.e., we made the so called {\it grey} approximation).
We solved a dimensionless version of Equation~\ref{eq:diffusion}: 
\begin{equation}
\frac{\partial u}{\partial t'}=\nabla\cdot(D'(z)\vec{\nabla}u)\;,
\end{equation}
where $t'=tD(h)/h^2$ and $D'(z)=D(z)/D(h)$. In this case the boundary conditions are 
\begin{equation}\label{boundary}
\lvert \lvert \vec{f}\rvert\rvert =\alpha vu\;\;\;\;,\;v=c/v_{d}\;,
\end{equation}
where $v_{d}=D(h)/h$. 
 A slab with a width $h=10^{16}\,\rm cm$, an opacity $\kappa=0.34$\,cm$^2$\,g$^{-1}$ and a constant mass density $\rho=1\times10^{-16}\,\rm g\,\rm cm^{-3}$ (corresponding e.g., to $3\,M_{\odot}$ of CSM in a slab with $L_x=L_y=8\,h$), corresponds to the unitless velocity $v\approx1$ and a diffusion time $t_d=h^23\kappa\rho/c=94$\,days. 
 
In order to minimize the effect of the finite size of the grid along $\vec{e_x}$ and $\vec{e_y}$, we took several precautions. We chose the dimensions of the grid along the $\vec{e_x}$ and $\vec{e_y}$ directions so that  $L_x=L_y$ and $L_x\gg h$. We checked that $L_x$ is large enough compared to $h$ so that a change in $L_x$ does not affect the results. Equation~\ref{boundary} only describes the boundary conditions along the $e_{\vec{z}}$ direction. We checked that we can apply reflective boundary conditions (i.e., $\vec{f}=\vec{0}$ at the boundaries) or absorbing boundary conditions (i.e., $u=0$ at the boundary) along the $\vec{e_x}$ and $\vec{e_y}$ directions, without affecting the results. We also checked for convergence of the code with respect to the time steps.


\subsection{Results}
\begin{figure}
\center
\includegraphics[scale=.50]{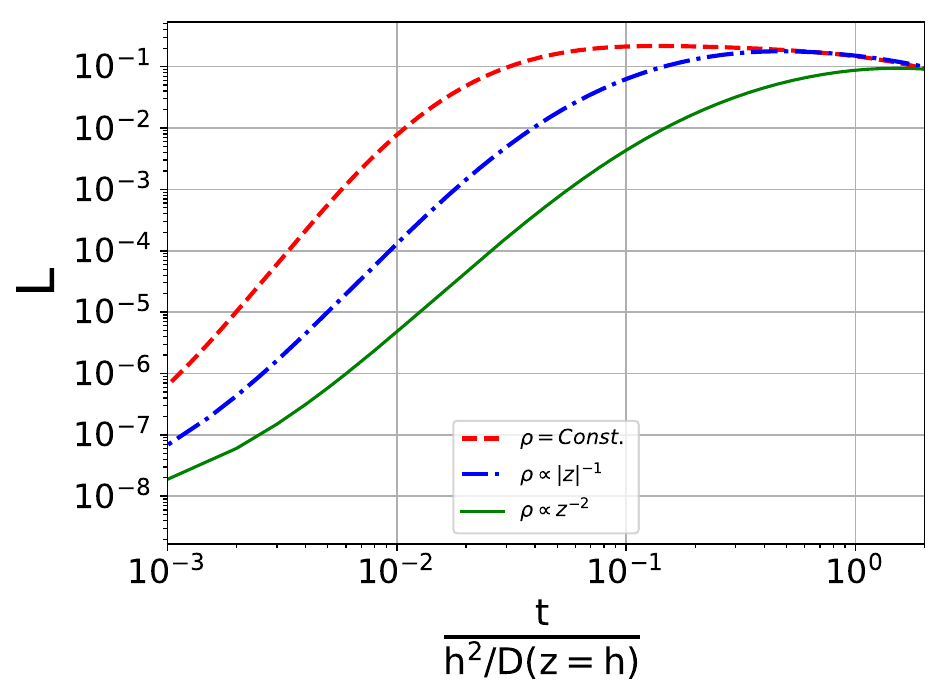}
\caption{Luminosity $L=\int_{S}f\,ds$ released at the surface of a slab with $v=1$, normalized so that the initial energy in the slab is 1. The luminosity is shown for a slab with a constant density profile (dashed line), a density profile $\rho\propto |z|^{-1}$ (dotted line) and a wind density profile $\rho\propto z^{-2}$ (continuous line).}
\label{fig:kernel}
\end{figure}

\begin{figure}
\center
\includegraphics[scale=.50]{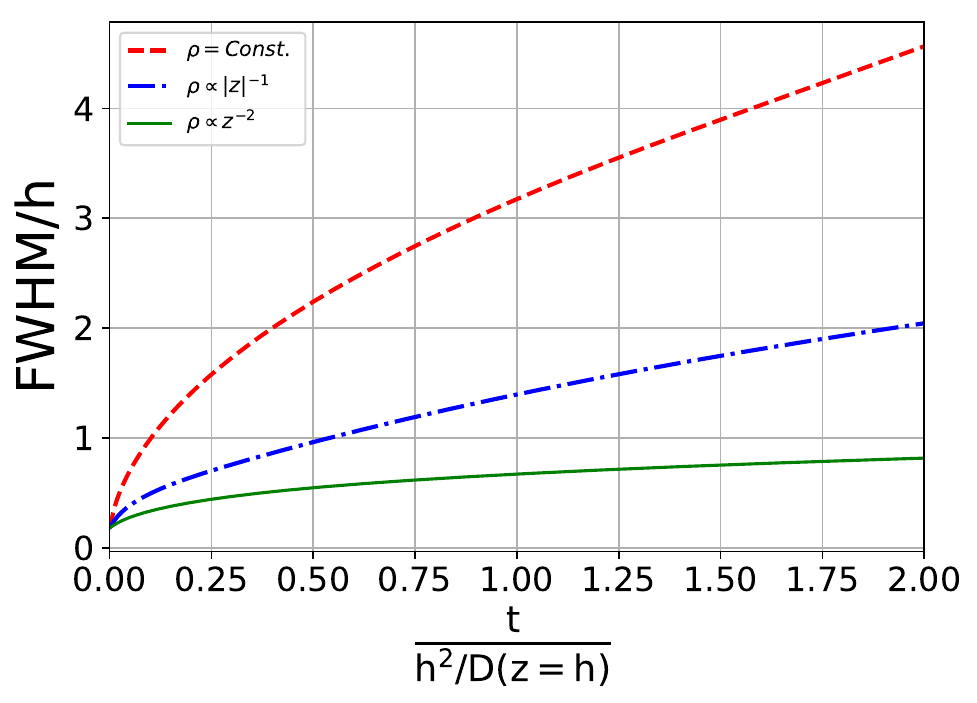}
\caption{Full width at half-maximum (FWHM) of the density of energy $u$ at the surface of a slab with $v=1$. The FWHM is used as a proxy for the black body radius measured by the observer. The FWHM is shown for a slab with a constant density profile (dashed line), a density profile $\rho\propto |z|^{-1}$ (dotted line) and a wind density profile $\rho\propto z^{-2}$ (continuous line).} 
\label{fig:radius}
\end{figure}

As photons diffuse in the slab, they reach the $z=h/2$ surface visible to the observer (see Figure~\ref{fig:cartoon}). In Figure~\ref{fig:kernel}, we show the evolution of the total flux of energy escaping from the surface of a slab with $v=1$, in response to an instantaneous deposition of energy at $t=0$. We use the full-width at half maximum (FWHM) of the energy density $u$ at the surface of the slab as a proxy for the radius seen by the observer (below we present another proxy for $r_{BB}$). In Figure~\ref{fig:radius}, we show that the FWHM grows in time for all three checked density profiles $\rho=Const.$, $\rho\propto |z|^{-1}$ and $\rho\propto z^{-2}$. We checked that varying the parameter $v$ does not change this qualitative result.

We would like to check whether our model can reproduce the decrease of the blackbody temperature $T_{BB}$ observed in Figure~\ref{fig:evo_param_TR}, in addition to producing a growing radius. Here again, given the simplicity of our model geometry, we are interested in the evolution of $T_{BB}$ rather than trying to fit its actual values. By modeling each cell of the $z=h/2$ surface as a blackbody with temperature $T\propto u(x,y)^{1/4}$ and summing up all the cell spectra, we can compute the overall spectrum of the surface. The resultant spectrum is well represented by a blackbody spectrum, which allows us to deduce the blackbody temperature of the surface. This strategy also provides an additional way to recover the growing radius, by using: 
\begin{equation}
L=\int_{S}f\,ds=\sigma T^4_{BB} 4\pi r_{BB}^2\;.
\end{equation}

\begin{figure}
\begin{center}
\includegraphics[scale=.40]{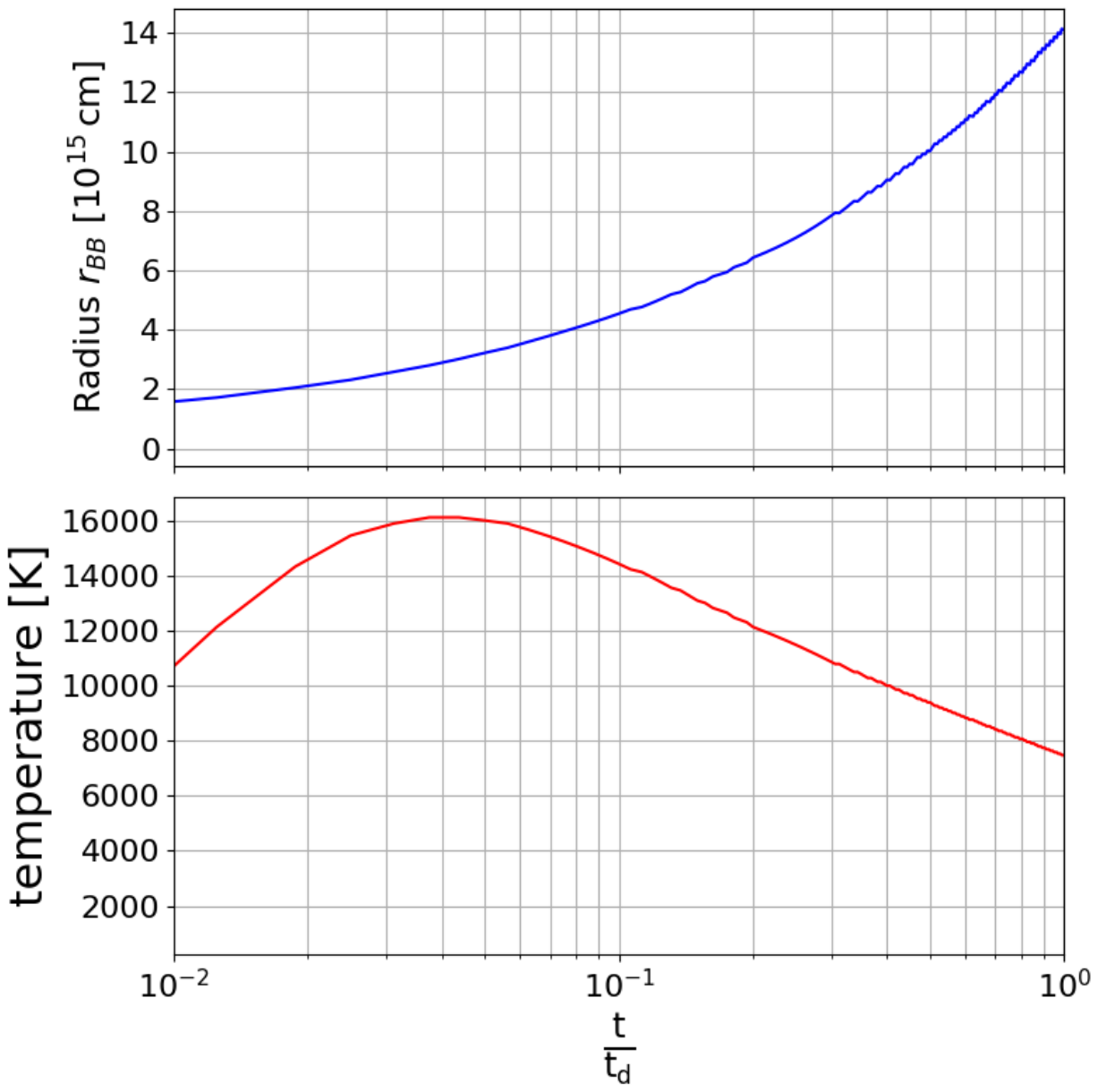}
\caption{The evolution in time of: (1) the blackbody radius $r_{BB}$ (top panel) and (2) the blackbody temperature (lower panel) at the surface of a slab with constant density. The diffusion equation was solved with an input energy $E_i=10^{51}$\,erg deposited in a slab of width $h=10^{16}\,\rm cm$ and constant mass density $\rho=1\times10^{-16}\,\rm g\,\rm cm^{-3}$, corresponding to $t_d=h^23\kappa\rho/c=94$\,days. The spectrum of the $z=h/2$ surface was deduced by summing the blackbody spectra of all the cells of the surface and $T_{BB}$ was then deduced by fitting a blackbody to the resultant spectrum. The blackbody radius $r_{BB}$ was deduced through the relation $r_{BB}=\sqrt{L/(4\pi\sigma T^4)}$. The aspherical geometry of the slab allows us to recover the increase of $r_{BB}$ and the decrease in $T_{BB}$.} 
\label{fig:temp}
\end{center}
\end{figure}

In Figure~\ref{fig:temp}, we show the evolution of the temperature $T_{BB}$ and radius $r_{BB}$, assuming $h=1\times10^{15}\,\rm cm$, a constant mass density $\rho=3\times10^{-14}\,\rm g\,\rm cm^{-3}$ and an input energy of $10^{51}$\,erg. Using this energy (calculated in Equation~\ref{eq:crude}) as the energy initially deposited in the slab, is equivalent to assuming that all the energy radiated by the SN explosion is released by the time it starts diffusing out through the CSM. This is only correct within the  assumption that the characteristic timescale for diffusion is much larger than the characteristic timescale for interaction between the ejecta and the CSM, which we showed is a correct assumption in the case of PTF\,12glz.
Figure~\ref{fig:temp} shows that the aspherical geometry of the slab allows to recover the increase of the radius $r_{BB}$ and the decrease of the temperature $T_{BB}$ observed in Figure~\ref{fig:evo_param_TR}.

\section{Conclusions}\label{sec:discussion}
We presented the observations of the supernova PTF\,12glz by the \textit{GALEX} space telescope and ground-based PTF. Radioactive decay is not sufficient to explain the decay of the light curve of PTF\,12glz and therefore other physical mechanisms must be involved. One possible - yet difficult to verify - scenario is that an internal engine powers the light curve. Another possible scenario - the standard explanation invoked in the case of Type IIn SNe, is that the light curve is powered by interaction between the ejecta and the CSM surrounding the SN. 


In the case of PTF\,12glz, the spectroscopic analysis is consistent with the following picture: at early times (two first spectra) both the ejecta and the shock are initially masked by a thick, slowly moving, photoionized CSM. At later times (two last spectra), the ejecta have emerged through - at least some of -  the optically thick layers and have reached CSM layers that are optically thin enough to expose the ejecta. CSM interaction may still play a role at late times, e.g., by heating the ejecta from the inside, and contributes to slowing the light curve decay.

The evolution of $r_{BB}$ - the radius of the deepest transparent emitting layer - seems to contradict this picture. At early times, i.e., at the very time when the opaque CSM seemingly obstructs our view of any growing structure, $r_{BB}$ grows by an order of magnitude, at a speed of $\sim8000$\,km\,s$^{-1}$. In addition to being inconsistent with the spectroscopic analysis, this is also in contradiction - to our knowledge - with all previous observations of either a constant or stalling blackbody radius in SNe IIn (as detailed in \S~\ref{sec:peculiar_r}).

If the bulk of the radiation from PTF\,12glz does come from interaction, the explanation for the growing blackbody radius may be geometrical. The question then is whether any peculiar structure of the CSM around the progenitor can reproduce the observations. In this work, we considered a simple aspherical structure of CSM: a slab. We modeled the radiation from an explosion embedded in a slab of CSM by numerically solving the radiative diffusion equation in a slab with different density profiles: $\rho=Const.$, $\rho\propto |z|^{-1}$ and a wind density profile $\rho\propto z^{-2}$. Although this model is simplistic, it allows recovery of the peculiar growth of the  blackbody radius $r_{BB}$ observed in the case of PTF\,12glz, as well as the decrease of its blackbody temperature $T_{BB}$. This configuration is not a unique geometrical solution and additional observations, e.g., of the polarization around PTF\,12glz would have been necessary to make it less speculative.

As new wide-field transient surveys such as the Zwicky Transient Facility (e.g., \citealt{Bellm2015,Laher2018}) are deployed, many more interacting SNe will be observed and quickly followed up with multiple-band observations. These may also be the brightest sources for the {\it ULTRASAT} $UV$ satellite mission \citep{Sagiv2014}. Some of these interacting SNe may exhibit the same peculiarities as PTF\,12glz. The methodology proposed in this paper offers a framework to analyze them. It could be elaborated upon, to model more complex aspherical geometries, e.g., $\eta$ Carinae-like shapes of the CSM, and give more quantitative predictions of the observables.

\acknowledgments
M.T.S. thanks Jonathan Morag, Adam Rubin, Yi Yang, Doron Kushnir, Anders Nyholm and Chalsea Harris for useful discussions. M.T.S. acknowledges support by a grant from IMOS/ISA, the Ilan Ramon fellowship from the Israel Ministry of Science and Technology and the Benoziyo center for Astrophysics at the Weizmann Institute of Science.

E.O.O is grateful for the support by grants from the Israel Science Foundation, Minerva, Israeli Ministry of Science, the US-Israel Binational Science Foundation, the Weizmann Institute and the I-CORE Program of the Planning and Budgeting Committee and the Israel Science Foundation.

A.G.-Y. is supported by the EU via ERC grant No. 725161, the Quantum Universe I-Core program, the ISF, the BSF Transformative program, IMOS via ISA and by a Kimmel award. 

This work is partly based on tools and data products produced by GAZPAR (\href{https://gazpar.lam.fr}{https://gazpar.lam.fr}) operated by CeSAM-LAM and IAP.



\bibliographystyle{apj} 
\bibliography{bibliograph.bib}


%

\end{document}